\documentclass[runningheads]{llncs}
\usepackage{graphicx}
\usepackage{xcolor}
\usepackage{url}
\begin{document}
\title{Visual deep learning-based explanation for neuritic plaques segmentation in Alzheimer's Disease using weakly annotated whole slide histopathological images}

\titlerunning{Visual DL-based explanation for neuritic plaques segmentation in Alzheimer's Disease}
\author{Gabriel Jimenez\inst{1}\orcidID{0000-0001-6474-6272} \and 
Anuradha Kar\inst{1}\orcidID{0000-0002-2543-1697} \and
Mehdi Ounissi\inst{1}\orcidID{0000-0003-0655-8112} \and
Léa Ingrassia\inst{1}\orcidID{0000-0002-4109-4870} \and
Susana Boluda\inst{1}\orcidID{0000-0002-8045-2706} \and
Benoît Delatour\inst{1}\orcidID{0000-0002-9910-9932} \and
Lev Stimmer\inst{1}\orcidID{0000-0003-2800-839X} \and
Daniel Racoceanu\inst{1}\orcidID{0000-0002-9416-1803}}

% index{Jimenez, Gabriel}
% index{Kar, Anuradha}
% index{Ounissi, Mehdi}
% index{Ingrassia, Léa}
% index{Boluda, Susana}
% index{Delatour, Benoît}
% index{Stimmer, Lev}
% index{Racoceanu, Daniel}

\authorrunning{G. Jimenez et al.}

\institute{Sorbonne Université, Paris Brain Institute - ICM, CNRS, Inria, Inserm, AP-HP, Hôpital de la Pitié Salpêtrière, Department of Neuropathology, DMU Neuroscience, F-75013, Paris, France \\ \email{daniel.racoceanu@sorbonne-universite.fr}}
\maketitle              % typeset the header of the contribution
\begin{abstract}
Quantifying the distribution and morphology of tau protein structures in brain tissues is key to diagnosing Alzheimer's Disease (AD) and its subtypes. Recently, deep learning (DL) models such as UNet have been successfully used for automatic segmentation of histopathological whole slide images (WSI) of biological tissues. In this study, we propose a DL-based methodology for semantic segmentation of tau lesions (i.e., neuritic plaques) in WSI of postmortem patients with AD. The state of the art in semantic segmentation of neuritic plaques in human WSI is very limited. Our study proposes a baseline able to generate a significant advantage for morphological analysis of these tauopathies for further stratification of AD patients. Essential discussions concerning biomarkers (ALZ50 versus AT8 tau antibodies), the imaging modality (different slide scanner resolutions), and the challenge of weak annotations are addressed within this seminal study. The analysis of the impact of context in plaque segmentation is important to understand the role of the micro-environment for reliable tau protein segmentation. In addition, by integrating visual interpretability, we are able to explain how the network focuses on a region of interest (ROI), giving additional insights to pathologists. Finally, the release of a new expert-annotated database and the code (\url{https://github.com/aramis-lab/miccai2022-stratifiad.git}) will be helpful for the scientific community to accelerate the development of new pipelines for human WSI processing in AD. 

\end{abstract}
\section{Introduction}
Accumulations of Amyloid-$\beta$ and tau protein aggregates, such as plaques in the brain gray matter, are well-known biomarkers of the neurodegenerative Alzheimer's disease (AD) \cite{ben1}. Quantitative estimation of plaques is typically done by pathologists manually or semi-automatically, using proprietary black-box software from histopathological images of the brain -- a time and effort-intensive process prone to human observation variability and errors. 
In recent times, deep learning (DL) based methods have shown promising results in digital pathology \cite{jano1} and incredibly high accuracy segmentation of digital whole slide images \cite{anant}. In \cite{wurtz}, three different DL models were used to segment tau aggregates (tangles) and nuclei in postmortem brain Whole Slide Images (WSIs). The three models included a fully convolutional neural network (FCN), UNet, and Segnet, the latter achieving the highest accuracy in terms of IoU. In \cite{signaevsky}, an FCN was trained on a dataset of 22 WSIs for semantic segmentation of tangle objects from postmortem brain WSIs. Their model can segment tangles of varying morphologies with high accuracy under diverse staining intensities. An FCN model was also used in \cite{Vega2021} to classify morphologies of tau protein aggregates in the gray and white matter regions from 37 WSIs representing multiple degenerative diseases. In \cite{manouskova2022}, tau aggregate analysis was done on a dataset of 6 WSIs with a combined classification-segmentation framework which achieved an F1 score of 81.3\% and 75.8\% on detection and segmentation tasks, respectively. 
Several domains in DL-based histopathological analysis of AD tauopathy remain unexplored. Firstly, most existing studies have used DL to segment tangles rather than plaques, which are harder to identify against the background gray matter due to their diffuse/sparse appearance. Secondly, annotations of whole slide images are frequently affected by errors by human annotators. In such cases, a  DL preliminary model may be trained using weakly annotated data and used to assist the expert in refining annotations. Thirdly, contemporary tau segmentation studies do not consider context information, which is essential in segmenting plaques from brain WSIs as these emerge as sparse objects against an extended background of gray matter. Finally, DL models with explainability features have not yet been applied in tau segmentation from WSIs. This is a critical requirement for DL models used in clinical applications \cite{explain1} \cite{Yamamoto2019}. The DL models should not only be able to identify regions of interest precisely but also give clinicians and general users the knowledge about which image features the model found necessary that influenced its decision. 
Based on the above, a DL pipeline for the segmentation of plaque regions in brain WSIs is presented in our study. This pipeline uses context and explainability features with a UNet-based semantic segmentation model to identify plaque features from WSIs.

\section{Methodology}
\label{sec:methodology}

\subsection{Dataset characteristics}
\label{sec:data_characteristics}
In this work, we analyzed eight whole slide images containing histological sections from the frontal cortices of patients with AD, which were provided by the French national brain biobank Neuro-CEB. Signed informed consent for autopsy and histologic analysis was obtained in all cases from the patients or their family members. The present cohort represents a common heterogeneity of AD cases, including slides with variable tau pathology (e.g., different object densities), variable staining quality, and variable tissue preservation. Sections of the frontal lobe were stained with AT8 antibody to reveal phosphorylated tau pathology, using a standardized immunohistochemistry protocol. Obtained slides were scanned using two Hamamatsu slide scanners (NanoZoomer 2.0-RS and NanoZoomer s60 with 227 nm/pixel and 221 nm/pixel resolution, respectively) at 40x initial magnification. The slides were used for human-CNN iterative object annotation resulting in about 4000 annotated and expert-validated Neuritic plaques. The labels, extracted in an XML format, constitute the segmentation ground truth.

\subsection{Data preparation}
\begin{figure}[!ht]
    \centering
    \includegraphics[scale=0.6] {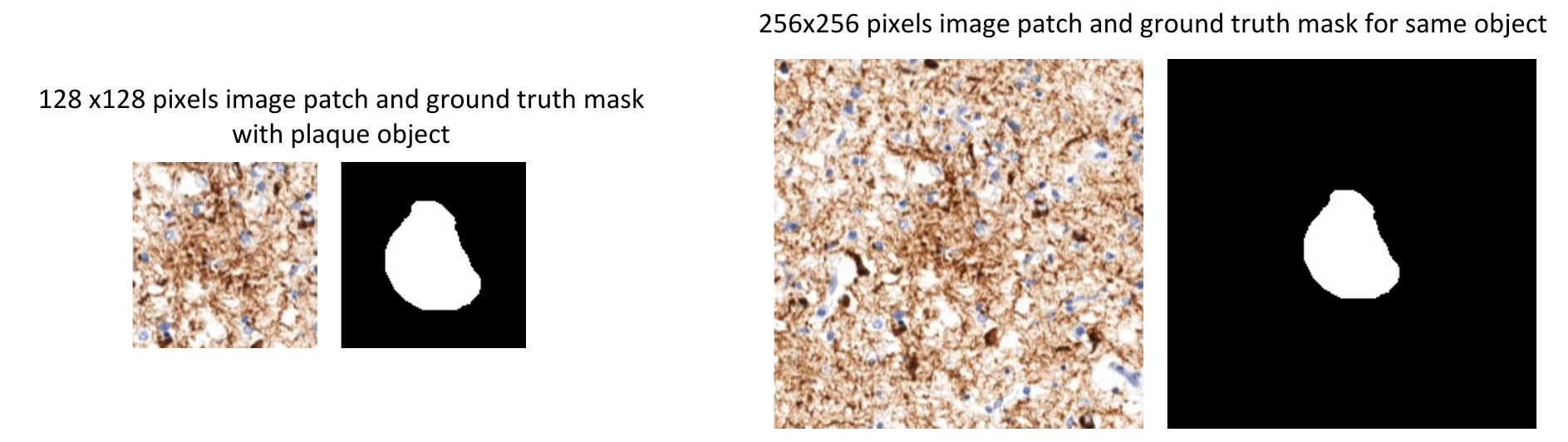}
    \caption{Example of plaque image for different levels of context.}
    \label{fig:context}
\end{figure}

From the WSIs, at 20x magnification, patches with two levels of context information were generated using an ROI-guided sampling method. The larger patches (256x256 pixels) capture a broader context containing object neighborhood and background pixels, whereas the smaller (128x 128 pixels) mainly focus on the plaque region without much context information. The amount of context present in each patch is quantified using a ratio of the area of annotated ROI to the total area of the patch. The plaque example in different patch sizes is shown in Fig~\ref{fig:context} (note that the bigger patch has additional objects-plaques). In addition, two different normalizations are used and compared: Macenko~\cite{macenko} and Vahadane~\cite{vahadane2015normalisation} methods.

\begin{figure}[!ht]
    \centering
    \includegraphics[scale=0.5] {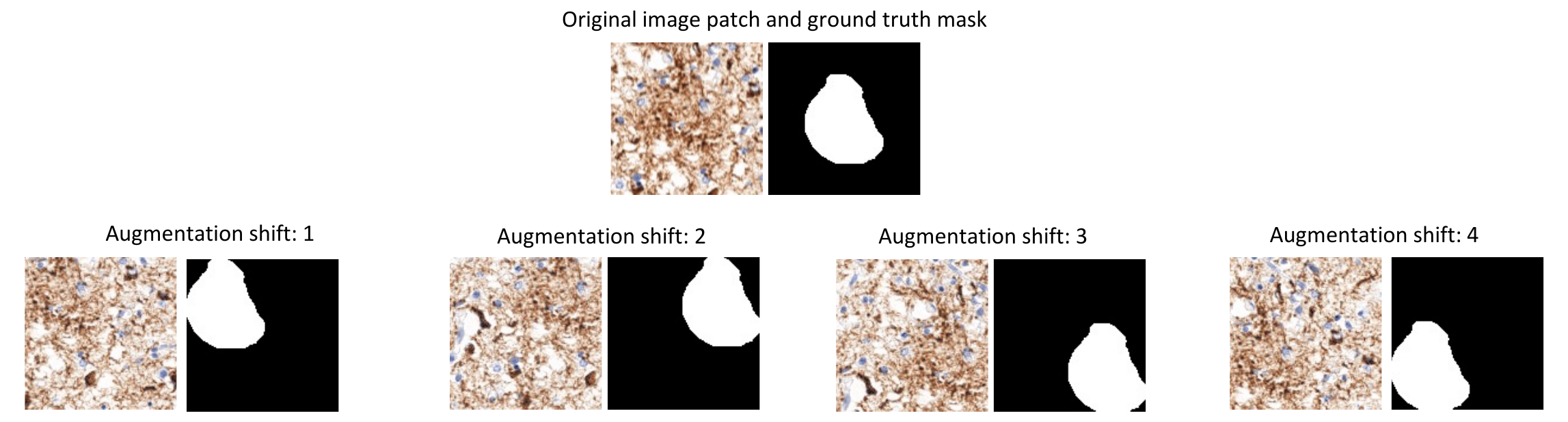}
    \caption{Example of ROI shifting augmentation.}
    \label{fig:ROI_aug}
\end{figure}

A new scheme for data augmentation was implemented based on ROI-shifting to prevent the networks' bias from focussing on the center location of plaques in the patches. Accordingly, the annotated plaque ROIs are shifted to four corners of a patch, producing a four-fold augmentation of each patch containing an object. This augmentation aims to train the UNet models robustly in the presence of variable neighborhood context information, especially when closely-spaced plaque objects are present. An example of this augmentation is shown in Fig~\ref{fig:ROI_aug}.

\subsection{Deep learning architecture for segmentation}

In order to segment the neuritic plaques, a UNet model adapted from \cite{Ronnenbunet} is used with modifications for accommodating context information within the WSI patches during training and testing. The model architecture is modified to work with the two datasets containing different patch sizes -- i.e., $128\times128$ (having low context information) and $256\times256$ pixels (having more information about the plaque neighborhood). For the first dataset, the UNet architecture consists of 3 downsampling and 3 upsampling convolutional blocks, in addition to the convolutional middle block. For the $256\times256$-patch-size dataset, we added a downsampling and upsampling convolutional block to the previous UNet model. For the downsampling block, we used a leaky ReLU activation function and ReLU for the upsampling block. In both blocks, we used batch-normalization following the suggestions in \cite{Ioffe2015BatchNA} and \cite{manouskova2022}. Dropout was used in each convolutional block with a probability of 0.5.

\subsection{Deep learning architecture for visual interpretation} 

In addition to the segmentation, we focus on deriving locations within the patches where the DL model found significant features from the plaque objects. Therefore, we used an attention UNet described in \cite{oktay2018attention}, which allows us to visualize the activated features at each iteration and evaluate qualitatively where the network focuses during training. The attention UNet architecture was also modified for the two different patch-size datasets following a configuration similar to the one described for the UNet.

\section{Experiments and results}
Data preparation and UNet experiments were executed on an 12-core Intel(R) Core i9-9920X @ 3.5GHz CPU with 128 GB RAM and two 12 GB RAM Nvidia GeForce RTX 2080 Ti GPUs. The attention UNet experiments run on a cluster (1 GPU Tesla V100S-PCIe-32GB, 12 CPU cores Intel(R) Xeon(R) Gold 6126 CPU @ 2.60GHz, and 80 GB of RAM). The average training and evaluation time of the UNet per epoch is approximately 2 minutes for the $128\times 128$ patch-size database and 5 minutes for the $256\times 256$ patch-size database. Meanwhile, for the attention UNet, approximately half the time is needed. On the other hand, data preprocessing takes 2 to 5 hours to process using parallel computation. Regarding memory consumption, we used at most 6 GB of GPU RAM for the larger patch dataset. In order to increase the performance, we cache the data and annotations first in CPU RAM and then move them to the GPU.

We randomly divided the 8 WSIs into 4 folds for the DL experiments. Then, we tested the network using a 4-fold cross-testing scheme, and with the remaining data from each test fold, we also performed a 3-fold cross-validation. In addition, we run a series of tests (using these folds) to select the loss function and the best optimizer for the UNet and attention UNet. We tested 4 loss functions (i.e., Focal loss, BCEwithLogits, Dice, and BCE-Dice loss) and 4 different optimizers (i.e., SGD, Adam, RMSProp, and AdaDelta). After the hyperparameter tuning, we obtained the best performance using the BCE-Dice loss with a 50\% balance between Dice and BCE (Binary Cross Entropy) and the Adadelta optimizer with $\rho = 0.9$ and a varying learning rate based on the evolution of the validation loss. Also, we implemented early stopping for training with a patience value of 15 epochs.

\subsection{Results from UNet architecture} % Gabriel, Anu
The segmentation evaluation metric used for all of the experiments regarding the UNet is the Dice score which is equivalent to the F1 score for binary segmentation problems. In the first experiment, the UNet model was trained with two datasets having different patch sizes: $128\times 128$ and $256\times 256$ pixels. The mean and standard deviations of the Dice coefficient for cross-validation and cross-testing are reported in Table~\ref{tab:dice_results1}. The patches were previously normalized using the Macenko method and then separated in their corresponding fold for training, validation, and testing following the scheme described above. We observe a decrease in the Dice score for larger patches having additional environmental context from the neuritic plaque.

\begin{table}[ht]
\centering
\caption{UNet results (Dice score) for 4-fold cross testing and 3-fold cross validation for different patch sizes.}
\begin{tabular}[t]{|c|c|c|c|}
\hline
Patch size & Normalization & Cross validation & Cross testing\\
\hline
$128\times128$ & Macenko & $ 0.6954 \pm 0.0289 $ & $0.6852 \pm 0.0260$\\
$256\times256$ & Macenko & $0.6600 \pm 0.0420 $ & $0.6460 \pm 0.0330$ \\
\hline
\end{tabular}
\label{tab:dice_results1}
\end{table}%

As described, the WSIs were acquired using two different scanners. Therefore, to study the impact of its properties, we divided the entire cohort into two independent datasets: 4 WSIs belonging to the NanoZoomer 2.0-RS and 4 WSIs scanned with the NanoZoomer s60. For both datasets, we only evaluate the performance of the DL architecture using 4-fold cross-validation and patches of $128\times 128$ pixels size. Additionally, we normalize each dataset independently (i.e., using two reference patches: one for the NanoZoomer 2.0-RS and one for the NanoZoomer s60) using the Macenko method. The Dice score obtained using the images from the higher resolution Hamamatsu NanoZoomer S60 scanner was $0.6345 \pm 0.0243$, whereas that from the NanoZoomer 2.0-RS was $0.7342 \pm 0.0063$.

We also study the effect of normalization in the entire dataset (8 WSIs). We normalized the patches from the $128\times 128$ dataset using Macenko and Vahadane methods, and we selected the best fold (i.e., highest Dice score in testing for the first experiment) to train, validate and test the UNet under different input color properties. Opposite to the results reported in~\cite{manouskova2022}, the Dice score obtained was higher using the Macenko method (0.7248 in testing) than the Vahadane (0.7098 in testing), even in validation (0.72313 for Macenko and 0.6864 for Vahadane). For a full list of results, see supplementary material.

\subsection{Visual deep learning interpretation} % Anu, Mehdi
The attention UNet model was trained using the $128\times 128$ and the $256\times 256$ patch size dataset, and the results are summarized in Table~\ref{tab:attunet_results1}. All images were normalized using the Macenko method, and we observed a similar trend as the UNet: better performance using patches containing less background information.

\begin{table}[ht]
\centering
\caption{Attention UNet results (Dice score) for 4-fold cross testing and 3-fold cross validation for different patch sizes.}
\begin{tabular}[t]{|c|c|c|c|}
\hline
Patch size & Normalization & Cross validation & Cross testing\\
\hline
$128\times128$ & Macenko & $ 0.7516 \pm 0.0334 $ & $0.6920 \pm 0.0254$\\
$256\times256$ & Macenko & $0.6931 \pm 0.0447 $ & $0.6342 \pm 0.0301$ \\
\hline
\end{tabular}
\label{tab:attunet_results1}
\end{table}%

\begin{figure}[!ht]
    \centering
    \includegraphics[scale=0.25] {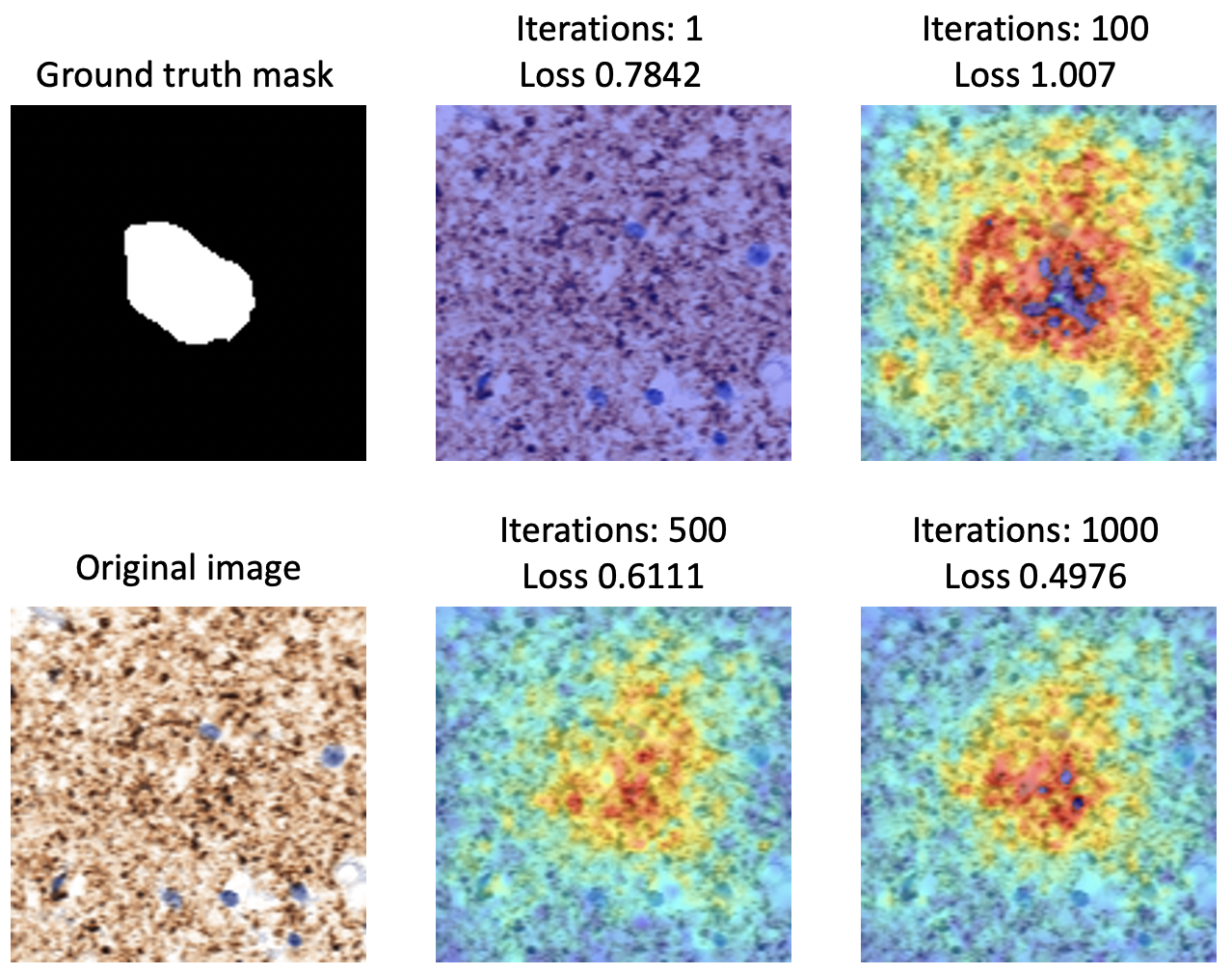}
    \caption{Global coherence of attention-UNet result with human annotation.}
    \label{fig:attunet1}
\end{figure}

An example segmentation result from the attention UNet model in a $128\times128$ patch containing a plaque object and its corresponding ground-truth mask is shown in Fig~\ref{fig:attunet1}. We observe that the attention UNet model finds significant activation features around the plaque object initially annotated by experts (see ground truth mask in Fig~\ref{fig:attunet1}). We also notice that the loss at iteration 100 increases over iteration 1; however, we clearly distinguish the region of the object (dark red color). After 1000 iterations, the loss decreases 50\% due to the fact that the Dice part of the BCE-Dice loss function influences the network into detecting a pattern very similar to the given ground truth.

\begin{figure}[!ht]
    \centering
    \includegraphics[scale=0.25] {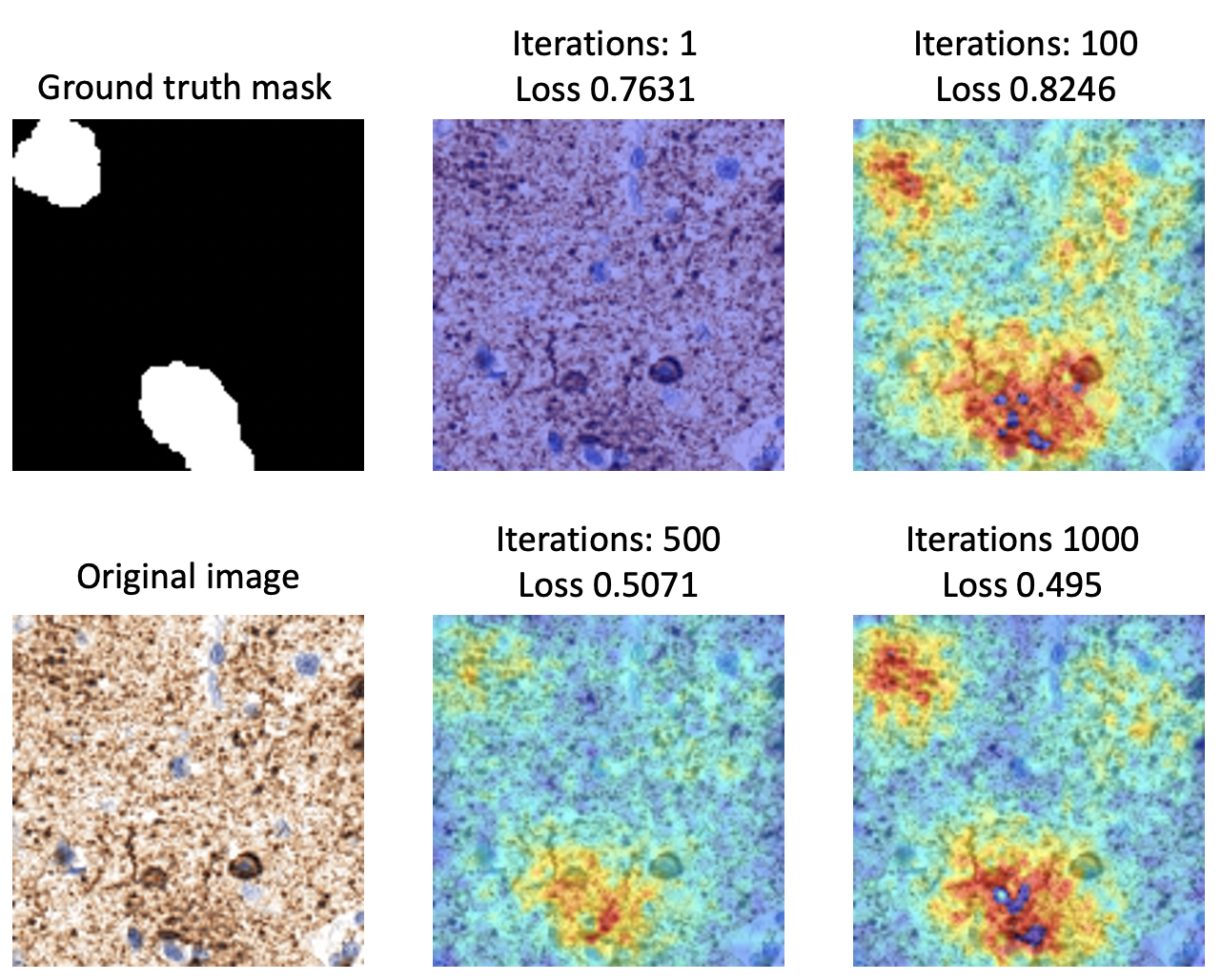}
    \caption{Focus progression using successive activation layers of attention-UNet.}
     \label{fig:attunet2}
\end{figure}

Another result from attention UNet is in Fig~\ref{fig:attunet2}. Here, the attention UNet focuses on 2 plaques initially annotated by a human expert. It also identifies strong activation features in regions with no ground truth annotations, which could indicate missed ROIs by human experts during the annotation process. Thus with the attention UNet, it is not only possible to segment the plaque objects but also to improve or refine the manual annotations by experts.

\begin{figure}[!ht]
    \centering
    \includegraphics[scale=0.7] {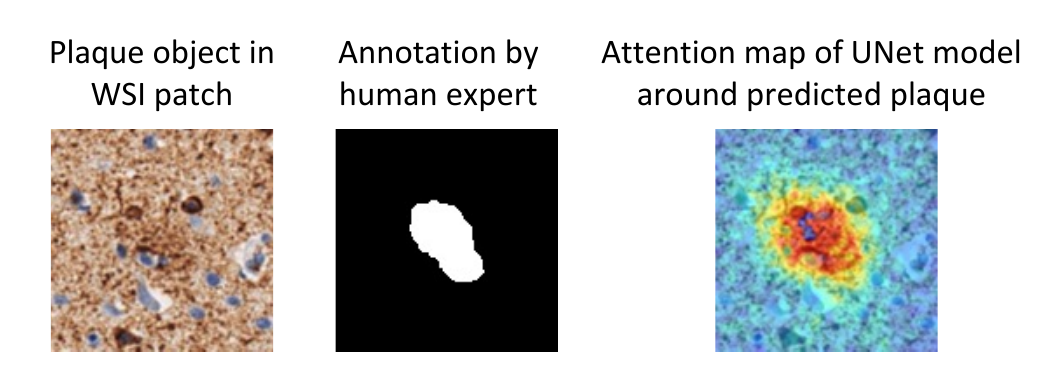}
    \caption{Improving human annotations using attention-based DL models.}
     \label{fig: attunet_vsexpert}
\end{figure}

Weak and imprecise annotations are frequently observed in histopathology arising from human or software errors. In such cases, deep learning attention maps could be useful to provide pathologists and biologists with refined annotations (e.g., precision on the boundaries of ROIs). An example is shown in Fig~\ref{fig: attunet_vsexpert} where DL attention maps are closer to the shape of actual ROIs compared to human-made annotations.

\section{Discussion and conclusion}
In the presented work, we studied/evaluated a number of factors that contribute to the segmentation of plaques from whole slide images using DL models. The key observations are the following: 
\begin{enumerate}
    \item Use of biomarkers: the study in~\cite{manouskova2022} uses the ALZ50 (used to discover compacted structures) biomarker, while our study uses the AT8 (majorly used in clinics, helps to discover all structures). We focus on AT8 in order to stay close to clinical protocols. The drawback is that this biomarker creates less compact structures meaning a slightly more difficult segmentation of the plaques, as our results support.
    \item Use of different modalities: using the AT8 biomarker, we analyzed 2 types of WSI scanners (see Section~\ref{sec:data_characteristics}) with different resolutions. High-resolution scanners amplify the annotation errors (human-software). Accordingly, some results concerning the high-resolution scanners have been affected, generating lower dice scores. 
    \item Context effect on results of DL models: We noticed that increasing the background information in the patches negatively affects the segmentation results, which can be explained by the imbalance between the foreground and background pixels. In future works, this will be addressed using adaptive loss functions to take advantage of context information around the ROIs.
    \item Attention maps: We observed that using the attention UNet model helps us see the weakness in the human-made annotations (see Fig~\ref{fig: attunet_vsexpert}), generating precious insights about the segmentation DL protocol, which can be used to refine the annotations by improving the border of the detected objects. These refined patterns can be used for a morphology and topology pipeline toward a robust AD patient's stratification proof. In addition, quantitative results show better performance of the same UNet architecture with attention blocks.
    \item Comparison with state-of-the-art commercial software: We compared our WSI segmentation results with those generated by a commercial software. This software uses a UNet architecture with a VGG encoder which is different from our model. Our system outperforms this software (Dice score 0.63 for test), using the same WSI as the ones used in this paper. Besides, in this software, neither information about how patches are generated nor the type of normalization or pre-processing perfomed on the dataset is available.
     
\end{enumerate}
Whole slide histopathology images whose sizes range in giga-pixels often contain thousands of objects per image. As seen for plaques in this study, it becomes more challenging when the objects being annotated do not have clear boundaries separating them from their surrounding environments, which may give rise to errors in human-made annotations. We saw an example of how DL models with visual explanation properties can help pathologists refine the ROI identification process. Our future challenge is to create deep learning assistive tools that can improve human-made few and weak annotations, a generic problem of a wide range of biomedical applications. 

\section*{Acknowlegements}
This research was supported by Mr Jean-Paul Baudecroux and The Big Brain Theory Program - Paris Brain Institute (ICM). The human samples were obtained from the Neuro-CEB brain bank (\url{https://www.neuroceb.org/en/}) (BRIF Number 0033-00011),  partly  funded  by  the  patients’ associations  ARSEP, ARSLA, “Connaître les Syndromes Cérébelleux”, France-DFT, France Parkinson and by Vaincre Alzheimer Fondation, to which we express our gratitude. We are also grateful to the patients and their families.

\bibliographystyle{splncs04}
\bibliography{references}

\end{document}

\title{Supplementary material - Paper ID: 2116}

\titlerunning{Visual DL-based explanation for neuritic plaques segmentation in Alzheimer's Disease}
\author{}

\authorrunning{G. Jimenez et al.}

\institute{}
\maketitle              % typeset the header of the contribution

\section{UNet and Attention UNet:}
% \usepackage{graphicx}
\begin{table}[htbp]
\centering
\caption{Architecture used: UNet. Patch size: 128x128 pixels (best fold is reported in bold font).}
\label{tab:my-table1}
\resizebox{\textwidth}{!}{%
\begin{tabular}{|c|c|c|c|c|c|c|c|c|}
\hline
{\color[HTML]{24292F} \textbf{fold\_name}} & {\color[HTML]{24292F} \textbf{dev\_dice}} & {\color[HTML]{24292F} \textbf{dev\_f1}} & {\color[HTML]{24292F} \textbf{dev\_recall}} & {\color[HTML]{24292F} \textbf{dev\_precision}} & {\color[HTML]{24292F} \textbf{test\_dice}} & {\color[HTML]{24292F} \textbf{test\_f1}} & {\color[HTML]{24292F} \textbf{test\_recall}} & {\color[HTML]{24292F} \textbf{test\_precision}} \\ \hline
{\color[HTML]{24292F} test\_00\_cv\_00} & {\color[HTML]{24292F} 0.7151} & {\color[HTML]{24292F} 0.7165} & {\color[HTML]{24292F} 0.6674} & {\color[HTML]{24292F} 0.8017} & {\color[HTML]{24292F} 0.6753} & {\color[HTML]{24292F} 0.6707} & {\color[HTML]{24292F} 0.668} & {\color[HTML]{24292F} 0.784} \\ \hline
{\color[HTML]{24292F} test\_00\_cv\_01} & {\color[HTML]{24292F} 0.7034} & {\color[HTML]{24292F} 0.6933} & {\color[HTML]{24292F} 0.699} & {\color[HTML]{24292F} 0.718} & {\color[HTML]{24292F} 0.7046} & {\color[HTML]{24292F} 0.7035} & {\color[HTML]{24292F} 0.7495} & {\color[HTML]{24292F} 0.7475} \\ \hline
{\color[HTML]{24292F} test\_00\_cv\_02} & {\color[HTML]{24292F} 0.6963} & {\color[HTML]{24292F} 0.6932} & {\color[HTML]{24292F} 0.6873} & {\color[HTML]{24292F} 0.7339} & {\color[HTML]{24292F} 0.6781} & {\color[HTML]{24292F} 0.6765} & {\color[HTML]{24292F} 0.7423} & {\color[HTML]{24292F} 0.7094} \\ \hline
{\color[HTML]{24292F} test\_01\_cv\_00} & {\color[HTML]{24292F} 0.7011} & {\color[HTML]{24292F} 0.7037} & {\color[HTML]{24292F} 0.714} & {\color[HTML]{24292F} 0.7239} & {\color[HTML]{24292F} 0.7032} & {\color[HTML]{24292F} 0.6962} & {\color[HTML]{24292F} 0.6684} & {\color[HTML]{24292F} 0.8052} \\ \hline
{\color[HTML]{24292F} \textbf{test\_01\_cv\_01}} & {\color[HTML]{24292F} \textbf{0.7231}} & {\color[HTML]{24292F} \textbf{0.7118}} & {\color[HTML]{24292F} \textbf{0.6763}} & {\color[HTML]{24292F} \textbf{0.7801}} & {\color[HTML]{24292F} \textbf{0.7248}} & {\color[HTML]{24292F} \textbf{0.7192}} & {\color[HTML]{24292F} \textbf{0.7105}} & {\color[HTML]{24292F} \textbf{0.8067}} \\ \hline
{\color[HTML]{24292F} test\_01\_cv\_02} & {\color[HTML]{24292F} 0.72} & {\color[HTML]{24292F} 0.7217} & {\color[HTML]{24292F} 0.7519} & {\color[HTML]{24292F} 0.7185} & {\color[HTML]{24292F} 0.7141} & {\color[HTML]{24292F} 0.7068} & {\color[HTML]{24292F} 0.6811} & {\color[HTML]{24292F} 0.8166} \\ \hline
{\color[HTML]{24292F} test\_02\_cv\_00} & {\color[HTML]{24292F} 0.709} & {\color[HTML]{24292F} 0.7156} & {\color[HTML]{24292F} 0.7195} & {\color[HTML]{24292F} 0.7423} & {\color[HTML]{24292F} 0.7027} & {\color[HTML]{24292F} 0.7004} & {\color[HTML]{24292F} 0.7043} & {\color[HTML]{24292F} 0.7855} \\ \hline
{\color[HTML]{24292F} test\_02\_cv\_01} & {\color[HTML]{24292F} 0.7127} & {\color[HTML]{24292F} 0.7195} & {\color[HTML]{24292F} 0.7012} & {\color[HTML]{24292F} 0.7618} & {\color[HTML]{24292F} 0.6643} & {\color[HTML]{24292F} 0.6608} & {\color[HTML]{24292F} 0.6444} & {\color[HTML]{24292F} 0.7996} \\ \hline
{\color[HTML]{24292F} test\_02\_cv\_02} & {\color[HTML]{24292F} 0.6807} & {\color[HTML]{24292F} 0.6838} & {\color[HTML]{24292F} 0.6862} & {\color[HTML]{24292F} 0.7167} & {\color[HTML]{24292F} 0.6306} & {\color[HTML]{24292F} 0.6296} & {\color[HTML]{24292F} 0.6634} & {\color[HTML]{24292F} 0.7316} \\ \hline
{\color[HTML]{24292F} test\_03\_cv\_00} & {\color[HTML]{24292F} 0.6765} & {\color[HTML]{24292F} 0.6813} & {\color[HTML]{24292F} 0.6788} & {\color[HTML]{24292F} 0.7183} & {\color[HTML]{24292F} 0.6845} & {\color[HTML]{24292F} 0.6855} & {\color[HTML]{24292F} 0.8061} & {\color[HTML]{24292F} 0.666} \\ \hline
{\color[HTML]{24292F} test\_03\_cv\_01} & {\color[HTML]{24292F} 0.6959} & {\color[HTML]{24292F} 0.6967} & {\color[HTML]{24292F} 0.6244} & {\color[HTML]{24292F} 0.8167} & {\color[HTML]{24292F} 0.6883} & {\color[HTML]{24292F} 0.879} & {\color[HTML]{24292F} 0.7981} & {\color[HTML]{24292F} 0.6777} \\ \hline
{\color[HTML]{24292F} test\_03\_cv\_02} & {\color[HTML]{24292F} 0.6112} & {\color[HTML]{24292F} 0.6008} & {\color[HTML]{24292F} 0.61} & {\color[HTML]{24292F} 0.6325} & {\color[HTML]{24292F} 0.6521} & {\color[HTML]{24292F} 0.6545} & {\color[HTML]{24292F} 0.8018} & {\color[HTML]{24292F} 0.6245} \\ \hline
{\color[HTML]{24292F} \textbf{mean}} & {\color[HTML]{24292F} 0.6954} & {\color[HTML]{24292F} 0.6948} & {\color[HTML]{24292F} 0.6847} & {\color[HTML]{24292F} 0.7387} & {\color[HTML]{24292F} 0.6852} & {\color[HTML]{24292F} 0.6986} & {\color[HTML]{24292F} 0.7198} & {\color[HTML]{24292F} 0.7462} \\ \hline
{\color[HTML]{24292F} \textbf{std}} & {\color[HTML]{24292F} 0.0289} & {\color[HTML]{24292F} 0.0313} & {\color[HTML]{24292F} 0.0373} & {\color[HTML]{24292F} 0.0462} & {\color[HTML]{24292F} 0.0260} & {\color[HTML]{24292F} 0.0596} & {\color[HTML]{24292F} 0.0561} & {\color[HTML]{24292F} 0.0615} \\ \hline
{\color[HTML]{24292F} \textbf{max}} & {\color[HTML]{24292F} 0.7231} & {\color[HTML]{24292F} 0.7217} & {\color[HTML]{24292F} 0.7519} & {\color[HTML]{24292F} 0.8167} & {\color[HTML]{24292F} 0.7248} & {\color[HTML]{24292F} 0.879} & {\color[HTML]{24292F} 0.8061} & {\color[HTML]{24292F} 0.8166} \\ \hline
{\color[HTML]{24292F} \textbf{min}} & {\color[HTML]{24292F} 0.6112} & {\color[HTML]{24292F} 0.6008} & {\color[HTML]{24292F} 0.61} & {\color[HTML]{24292F} 0.6325} & {\color[HTML]{24292F} 0.6306} & {\color[HTML]{24292F} 0.6296} & {\color[HTML]{24292F} 0.6444} & {\color[HTML]{24292F} 0.6245} \\ \hline
\end{tabular}%
}
\end{table}

% \usepackage{graphicx}
\begin{table}[htbp]
\centering
\caption{Architecture used: Attention UNet. Patch size: 128x128 pixels (best fold is reported in bold font).}
\label{tab:my-table2}
\resizebox{\textwidth}{!}{%
\begin{tabular}{|c|c|c|c|c|c|c|c|c|}
\hline
{\color[HTML]{24292F} \textbf{fold\_name}} & {\color[HTML]{24292F} \textbf{dev\_dice}} & {\color[HTML]{24292F} \textbf{dev\_f1}} & {\color[HTML]{24292F} \textbf{dev\_recall}} & {\color[HTML]{24292F} \textbf{dev\_precision}} & {\color[HTML]{24292F} \textbf{test\_dice}} & {\color[HTML]{24292F} \textbf{test\_f1}} & {\color[HTML]{24292F} \textbf{test\_recall}} & {\color[HTML]{24292F} \textbf{test\_precision}} \\ \hline
{\color[HTML]{24292F} test\_00\_cv\_00} & {\color[HTML]{24292F} 0.7654} & {\color[HTML]{24292F} 0.7654} & {\color[HTML]{24292F} 0.7833} & {\color[HTML]{24292F} 0.8000} & {\color[HTML]{24292F} 0.6405} & {\color[HTML]{24292F} 0.6405} & {\color[HTML]{24292F} 0.5959} & {\color[HTML]{24292F} 0.8177} \\ \hline
{\color[HTML]{24292F} test\_00\_cv\_01} & {\color[HTML]{24292F} 0.7216} & {\color[HTML]{24292F} 0.7216} & {\color[HTML]{24292F} 0.7293} & {\color[HTML]{24292F} 0.7890} & {\color[HTML]{24292F} 0.6734} & {\color[HTML]{24292F} 0.6734} & {\color[HTML]{24292F} 0.6387} & {\color[HTML]{24292F} 0.8258} \\ \hline
{\color[HTML]{24292F} test\_00\_cv\_02} & {\color[HTML]{24292F} 0.7428} & {\color[HTML]{24292F} 0.7422} & {\color[HTML]{24292F} 0.7425} & {\color[HTML]{24292F} 0.8108} & {\color[HTML]{24292F} 0.6499} & {\color[HTML]{24292F} 0.6499} & {\color[HTML]{24292F} 0.6320} & {\color[HTML]{24292F} 0.7883} \\ \hline
{\color[HTML]{24292F} test\_01\_cv\_00} & {\color[HTML]{24292F} 0.7646} & {\color[HTML]{24292F} 0.7646} & {\color[HTML]{24292F} 0.7876} & {\color[HTML]{24292F} 0.7934} & {\color[HTML]{24292F} 0.6851} & {\color[HTML]{24292F} 0.6851} & {\color[HTML]{24292F} 0.6582} & {\color[HTML]{24292F} 0.7979} \\ \hline
{\color[HTML]{24292F} test\_01\_cv\_01} & {\color[HTML]{24292F} 0.7272} & {\color[HTML]{24292F} 0.7272} & {\color[HTML]{24292F} 0.7916} & {\color[HTML]{24292F} 0.7299} & {\color[HTML]{24292F} 0.7128} & {\color[HTML]{24292F} 0.7122} & {\color[HTML]{24292F} 0.7454} & {\color[HTML]{24292F} 0.7522} \\ \hline
{\color[HTML]{24292F} test\_01\_cv\_02} & {\color[HTML]{24292F} 0.7335} & {\color[HTML]{24292F} 0.7335} & {\color[HTML]{24292F} 0.7593} & {\color[HTML]{24292F} 0.7772} & {\color[HTML]{24292F} 0.6909} & {\color[HTML]{24292F} 0.6909} & {\color[HTML]{24292F} 0.6836} & {\color[HTML]{24292F} 0.7890} \\ \hline
{\color[HTML]{24292F} test\_02\_cv\_00} & {\color[HTML]{24292F} 0.7658} & {\color[HTML]{24292F} 0.7658} & {\color[HTML]{24292F} 0.7946} & {\color[HTML]{24292F} 0.7900} & {\color[HTML]{24292F} 0.7184} & {\color[HTML]{24292F} 0.7184} & {\color[HTML]{24292F} 0.7109} & {\color[HTML]{24292F} 0.8060} \\ \hline
{\color[HTML]{24292F} \textbf{test\_02\_cv\_01}} & {\color[HTML]{24292F} \textbf{0.7159}} & {\color[HTML]{24292F} \textbf{0.7153}} & {\color[HTML]{24292F} \textbf{0.7565}} & {\color[HTML]{24292F} \textbf{0.7429}} & {\color[HTML]{24292F} \textbf{0.7263}} & {\color[HTML]{24292F} \textbf{0.7263}} & {\color[HTML]{24292F} \textbf{0.7950}} & {\color[HTML]{24292F} \textbf{0.7254}} \\ \hline
{\color[HTML]{24292F} test\_02\_cv\_02} & {\color[HTML]{24292F} 0.7809} & {\color[HTML]{24292F} 0.7809} & {\color[HTML]{24292F} 0.8232} & {\color[HTML]{24292F} 0.7855} & {\color[HTML]{24292F} 0.6948} & {\color[HTML]{24292F} 0.6948} & {\color[HTML]{24292F} 0.7241} & {\color[HTML]{24292F} 0.7488} \\ \hline
{\color[HTML]{24292F} test\_03\_cv\_00} & {\color[HTML]{24292F} 0.8193} & {\color[HTML]{24292F} 0.8193} & {\color[HTML]{24292F} 0.8180} & {\color[HTML]{24292F} 0.8559} & {\color[HTML]{24292F} 0.6959} & {\color[HTML]{24292F} 0.6959} & {\color[HTML]{24292F} 0.6839} & {\color[HTML]{24292F} 0.8008} \\ \hline
{\color[HTML]{24292F} test\_03\_cv\_01} & {\color[HTML]{24292F} 0.6962} & {\color[HTML]{24292F} 0.6962} & {\color[HTML]{24292F} 0.6843} & {\color[HTML]{24292F} 0.7908} & {\color[HTML]{24292F} 0.7140} & {\color[HTML]{24292F} 0.7140} & {\color[HTML]{24292F} 0.7781} & {\color[HTML]{24292F} 0.7281} \\ \hline
{\color[HTML]{24292F} test\_03\_cv\_02} & {\color[HTML]{24292F} 0.7857} & {\color[HTML]{24292F} 0.7857} & {\color[HTML]{24292F} 0.8297} & {\color[HTML]{24292F} 0.7896} & {\color[HTML]{24292F} 0.7024} & {\color[HTML]{24292F} 0.7024} & {\color[HTML]{24292F} 0.8435} & {\color[HTML]{24292F} 0.6571} \\ \hline
{\color[HTML]{24292F} \textbf{mean}} & {\color[HTML]{24292F} 0.7516} & {\color[HTML]{24292F} 0.7515} & {\color[HTML]{24292F} 0.7750} & {\color[HTML]{24292F} 0.7879} & {\color[HTML]{24292F} 0.6920} & {\color[HTML]{24292F} 0.6920} & {\color[HTML]{24292F} 0.7074} & {\color[HTML]{24292F} 0.7698} \\ \hline
{\color[HTML]{24292F} \textbf{std}} & {\color[HTML]{24292F} 0.0334} & {\color[HTML]{24292F} 0.0335} & {\color[HTML]{24292F} 0.0408} & {\color[HTML]{24292F} 0.0301} & {\color[HTML]{24292F} 0.0254} & {\color[HTML]{24292F} 0.0254} & {\color[HTML]{24292F} 0.0703} & {\color[HTML]{24292F} 0.0469} \\ \hline
{\color[HTML]{24292F} \textbf{max}} & {\color[HTML]{24292F} 0.8193} & {\color[HTML]{24292F} 0.8193} & {\color[HTML]{24292F} 0.8297} & {\color[HTML]{24292F} 0.8559} & {\color[HTML]{24292F} 0.7263} & {\color[HTML]{24292F} 0.7263} & {\color[HTML]{24292F} 0.8435} & {\color[HTML]{24292F} 0.8258} \\ \hline
{\color[HTML]{24292F} \textbf{min}} & {\color[HTML]{24292F} 0.6962} & {\color[HTML]{24292F} 0.6962} & {\color[HTML]{24292F} 0.6843} & {\color[HTML]{24292F} 0.7299} & {\color[HTML]{24292F} 0.6405} & {\color[HTML]{24292F} 0.6405} & {\color[HTML]{24292F} 0.5959} & {\color[HTML]{24292F} 0.6571} \\ \hline
\end{tabular}%
}
\end{table}

% \usepackage{graphicx}
\begin{table}[htbp]
\centering
\caption{Architecture used: UNet. Patch size: 256x256 pixels (best fold is reported in bold font).}
\label{tab:my-table3}
\resizebox{\textwidth}{!}{%
\begin{tabular}{|c|c|c|c|c|c|c|c|c|}
\hline
{\color[HTML]{24292F} \textbf{fold\_name}} & {\color[HTML]{24292F} \textbf{dev\_dice}} & {\color[HTML]{24292F} \textbf{dev\_f1}} & {\color[HTML]{24292F} \textbf{dev\_recall}} & {\color[HTML]{24292F} \textbf{dev\_precision}} & {\color[HTML]{24292F} \textbf{test\_dice}} & {\color[HTML]{24292F} \textbf{test\_f1}} & {\color[HTML]{24292F} \textbf{test\_recall}} & {\color[HTML]{24292F} \textbf{test\_precision}} \\ \hline
{\color[HTML]{24292F} test\_00\_cv\_00} & {\color[HTML]{24292F} 0.707} & {\color[HTML]{24292F} 0.7095} & {\color[HTML]{24292F} 0.6557} & {\color[HTML]{24292F} 0.7962} & {\color[HTML]{24292F} 0.6423} & {\color[HTML]{24292F} 0.6363} & {\color[HTML]{24292F} 0.5839} & {\color[HTML]{24292F} 0.8071} \\ \hline
{\color[HTML]{24292F} test\_00\_cv\_01} & {\color[HTML]{24292F} 0.6509} & {\color[HTML]{24292F} 0.6563} & {\color[HTML]{24292F} 0.6744} & {\color[HTML]{24292F} 0.6604} & {\color[HTML]{24292F} 0.6783} & {\color[HTML]{24292F} 0.6743} & {\color[HTML]{24292F} 0.6563} & {\color[HTML]{24292F} 0.7767} \\ \hline
{\color[HTML]{24292F} test\_00\_cv\_02} & {\color[HTML]{24292F} 0.6663} & {\color[HTML]{24292F} 0.6657} & {\color[HTML]{24292F} 0.6725} & {\color[HTML]{24292F} 0.6826} & {\color[HTML]{24292F} 0.6368} & {\color[HTML]{24292F} 0.6322} & {\color[HTML]{24292F} 0.6356} & {\color[HTML]{24292F} 0.7269} \\ \hline
{\color[HTML]{24292F} \textbf{test\_01\_cv\_00}} & {\color[HTML]{24292F} \textbf{0.6939}} & {\color[HTML]{24292F} \textbf{0.7099}} & {\color[HTML]{24292F} \textbf{0.7041}} & {\color[HTML]{24292F} \textbf{0.7319}} & {\color[HTML]{24292F} \textbf{0.6963}} & {\color[HTML]{24292F} \textbf{0.6893}} & {\color[HTML]{24292F} \textbf{0.6632}} & {\color[HTML]{24292F} \textbf{0.7998}} \\ \hline
{\color[HTML]{24292F} test\_01\_cv\_01} & {\color[HTML]{24292F} 0.6411} & {\color[HTML]{24292F} 0.642} & {\color[HTML]{24292F} 0.6018} & {\color[HTML]{24292F} 0.7157} & {\color[HTML]{24292F} 0.6856} & {\color[HTML]{24292F} 0.6785} & {\color[HTML]{24292F} 0.6468} & {\color[HTML]{24292F} 0.8008} \\ \hline
{\color[HTML]{24292F} test\_01\_cv\_02} & {\color[HTML]{24292F} 0.6793} & {\color[HTML]{24292F} 0.6829} & {\color[HTML]{24292F} 0.6892} & {\color[HTML]{24292F} 0.6979} & {\color[HTML]{24292F} 0.6718} & {\color[HTML]{24292F} 0.6654} & {\color[HTML]{24292F} 0.6487} & {\color[HTML]{24292F} 0.7732} \\ \hline
{\color[HTML]{24292F} test\_02\_cv\_00} & {\color[HTML]{24292F} 0.6893} & {\color[HTML]{24292F} 0.7077} & {\color[HTML]{24292F} 0.716} & {\color[HTML]{24292F} 0.716} & {\color[HTML]{24292F} 0.6395} & {\color[HTML]{24292F} 0.6352} & {\color[HTML]{24292F} 0.6104} & {\color[HTML]{24292F} 0.766} \\ \hline
{\color[HTML]{24292F} test\_02\_cv\_01} & {\color[HTML]{24292F} 0.6911} & {\color[HTML]{24292F} 0.698} & {\color[HTML]{24292F} 0.6344} & {\color[HTML]{24292F} 0.8027} & {\color[HTML]{24292F} 0.621} & {\color[HTML]{24292F} 0.6155} & {\color[HTML]{24292F} 0.5748} & {\color[HTML]{24292F} 0.7816} \\ \hline
{\color[HTML]{24292F} test\_02\_cv\_02} & {\color[HTML]{24292F} 0.6462} & {\color[HTML]{24292F} 0.649} & {\color[HTML]{24292F} 0.6593} & {\color[HTML]{24292F} 0.6638} & {\color[HTML]{24292F} 0.5949} & {\color[HTML]{24292F} 0.595} & {\color[HTML]{24292F} 0.6676} & {\color[HTML]{24292F} 0.6185} \\ \hline
{\color[HTML]{24292F} test\_03\_cv\_00} & {\color[HTML]{24292F} 0.6432} & {\color[HTML]{24292F} 0.6529} & {\color[HTML]{24292F} 0.6323} & {\color[HTML]{24292F} 0.7029} & {\color[HTML]{24292F} 0.6507} & {\color[HTML]{24292F} 0.65} & {\color[HTML]{24292F} 0.7633} & {\color[HTML]{24292F} 0.6242} \\ \hline
{\color[HTML]{24292F} test\_03\_cv\_01} & {\color[HTML]{24292F} 0.6747} & {\color[HTML]{24292F} 0.6877} & {\color[HTML]{24292F} 0.6581} & {\color[HTML]{24292F} 0.7453} & {\color[HTML]{24292F} 0.6507} & {\color[HTML]{24292F} 0.6521} & {\color[HTML]{24292F} 0.8146} & {\color[HTML]{24292F} 0.5945} \\ \hline
{\color[HTML]{24292F} test\_03\_cv\_02} & {\color[HTML]{24292F} 0.5391} & {\color[HTML]{24292F} 0.5355} & {\color[HTML]{24292F} 0.5558} & {\color[HTML]{24292F} 0.5535} & {\color[HTML]{24292F} 0.5834} & {\color[HTML]{24292F} 0.5856} & {\color[HTML]{24292F} 0.8118} & {\color[HTML]{24292F} 0.5149} \\ \hline
{\color[HTML]{24292F} \textbf{mean}} & {\color[HTML]{24292F} 0.660} & {\color[HTML]{24292F} 0.666} & {\color[HTML]{24292F} 0.654} & {\color[HTML]{24292F} 0.706} & {\color[HTML]{24292F} 0.646} & {\color[HTML]{24292F} 0.642} & {\color[HTML]{24292F} 0.673} & {\color[HTML]{24292F} 0.715} \\ \hline
{\color[HTML]{24292F} \textbf{std}} & {\color[HTML]{24292F} 0.042} & {\color[HTML]{24292F} 0.046} & {\color[HTML]{24292F} 0.042} & {\color[HTML]{24292F} 0.063} & {\color[HTML]{24292F} 0.033} & {\color[HTML]{24292F} 0.031} & {\color[HTML]{24292F} 0.077} & {\color[HTML]{24292F} 0.096} \\ \hline
{\color[HTML]{24292F} \textbf{max}} & {\color[HTML]{24292F} 0.707} & {\color[HTML]{24292F} 0.710} & {\color[HTML]{24292F} 0.716} & {\color[HTML]{24292F} 0.803} & {\color[HTML]{24292F} 0.696} & {\color[HTML]{24292F} 0.689} & {\color[HTML]{24292F} 0.815} & {\color[HTML]{24292F} 0.807} \\ \hline
{\color[HTML]{24292F} \textbf{min}} & {\color[HTML]{24292F} 0.5391} & {\color[HTML]{24292F} 0.5355} & {\color[HTML]{24292F} 0.5558} & {\color[HTML]{24292F} 0.5535} & {\color[HTML]{24292F} 0.5834} & {\color[HTML]{24292F} 0.5856} & {\color[HTML]{24292F} 0.5748} & {\color[HTML]{24292F} 0.5149} \\ \hline
\end{tabular}%
}
\end{table}

% \usepackage{graphicx}
\begin{table}[htbp]
\centering
\caption{Architecture used: Attention UNet. Patch size: 256x256 pixels (best fold is reported in bold font).}
\label{tab:my-table4}
\resizebox{\textwidth}{!}{%
\begin{tabular}{|c|c|c|c|c|c|c|c|c|}
\hline
{\color[HTML]{24292F} \textbf{fold\_name}} & {\color[HTML]{24292F} \textbf{dev\_dice}} & {\color[HTML]{24292F} \textbf{dev\_f1}} & {\color[HTML]{24292F} \textbf{dev\_recall}} & {\color[HTML]{24292F} \textbf{dev\_precision}} & {\color[HTML]{24292F} \textbf{test\_dice}} & {\color[HTML]{24292F} \textbf{test\_f1}} & {\color[HTML]{24292F} \textbf{test\_recall}} & {\color[HTML]{24292F} \textbf{test\_precision}} \\ \hline
{\color[HTML]{24292F} test\_00\_cv\_00} & {\color[HTML]{24292F} 0.7288} & {\color[HTML]{24292F} 0.7288} & {\color[HTML]{24292F} 0.8318} & {\color[HTML]{24292F} 0.6819} & {\color[HTML]{24292F} 0.6439} & {\color[HTML]{24292F} 0.6439} & {\color[HTML]{24292F} 0.6312} & {\color[HTML]{24292F} 0.7429} \\ \hline
{\color[HTML]{24292F} test\_00\_cv\_01} & {\color[HTML]{24292F} 0.6698} & {\color[HTML]{24292F} 0.6698} & {\color[HTML]{24292F} 0.7152} & {\color[HTML]{24292F} 0.6855} & {\color[HTML]{24292F} 0.6038} & {\color[HTML]{24292F} 0.6038} & {\color[HTML]{24292F} 0.5860} & {\color[HTML]{24292F} 0.7462} \\ \hline
{\color[HTML]{24292F} test\_00\_cv\_02} & {\color[HTML]{24292F} 0.6644} & {\color[HTML]{24292F} 0.6644} & {\color[HTML]{24292F} 0.6396} & {\color[HTML]{24292F} 0.7920} & {\color[HTML]{24292F} 0.5638} & {\color[HTML]{24292F} 0.5638} & {\color[HTML]{24292F} 0.5132} & {\color[HTML]{24292F} 0.7701} \\ \hline
{\color[HTML]{24292F} test\_01\_cv\_00} & {\color[HTML]{24292F} 0.7349} & {\color[HTML]{24292F} 0.7349} & {\color[HTML]{24292F} 0.8246} & {\color[HTML]{24292F} 0.6971} & {\color[HTML]{24292F} 0.6560} & {\color[HTML]{24292F} 0.6560} & {\color[HTML]{24292F} 0.6724} & {\color[HTML]{24292F} 0.7162} \\ \hline
{\color[HTML]{24292F} \textbf{test\_01\_cv\_01}} & {\color[HTML]{24292F} \textbf{0.6471}} & {\color[HTML]{24292F} \textbf{0.6471}} & {\color[HTML]{24292F} \textbf{0.6618}} & {\color[HTML]{24292F} \textbf{0.7102}} & {\color[HTML]{24292F} \textbf{0.6796}} & {\color[HTML]{24292F} \textbf{0.6790}} & {\color[HTML]{24292F} \textbf{0.6746}} & {\color[HTML]{24292F} \textbf{0.7599}} \\ \hline
{\color[HTML]{24292F} test\_01\_cv\_02} & {\color[HTML]{24292F} 0.6774} & {\color[HTML]{24292F} 0.6774} & {\color[HTML]{24292F} 0.6539} & {\color[HTML]{24292F} 0.7842} & {\color[HTML]{24292F} 0.6307} & {\color[HTML]{24292F} 0.6307} & {\color[HTML]{24292F} 0.5998} & {\color[HTML]{24292F} 0.7847} \\ \hline
{\color[HTML]{24292F} test\_02\_cv\_00} & {\color[HTML]{24292F} 0.7463} & {\color[HTML]{24292F} 0.7463} & {\color[HTML]{24292F} 0.8221} & {\color[HTML]{24292F} 0.7172} & {\color[HTML]{24292F} 0.6323} & {\color[HTML]{24292F} 0.6323} & {\color[HTML]{24292F} 0.6289} & {\color[HTML]{24292F} 0.7087} \\ \hline
{\color[HTML]{24292F} test\_02\_cv\_01} & {\color[HTML]{24292F} 0.6767} & {\color[HTML]{24292F} 0.6761} & {\color[HTML]{24292F} 0.7026} & {\color[HTML]{24292F} 0.7199} & {\color[HTML]{24292F} 0.6497} & {\color[HTML]{24292F} 0.6497} & {\color[HTML]{24292F} 0.6715} & {\color[HTML]{24292F} 0.6983} \\ \hline
{\color[HTML]{24292F} test\_02\_cv\_02} & {\color[HTML]{24292F} 0.6883} & {\color[HTML]{24292F} 0.6883} & {\color[HTML]{24292F} 0.7973} & {\color[HTML]{24292F} 0.6619} & {\color[HTML]{24292F} 0.6194} & {\color[HTML]{24292F} 0.6194} & {\color[HTML]{24292F} 0.6652} & {\color[HTML]{24292F} 0.6538} \\ \hline
{\color[HTML]{24292F} test\_03\_cv\_00} & {\color[HTML]{24292F} 0.7855} & {\color[HTML]{24292F} 0.7855} & {\color[HTML]{24292F} 0.7662} & {\color[HTML]{24292F} 0.8413} & {\color[HTML]{24292F} 0.6413} & {\color[HTML]{24292F} 0.6413} & {\color[HTML]{24292F} 0.6032} & {\color[HTML]{24292F} 0.7821} \\ \hline
{\color[HTML]{24292F} test\_03\_cv\_01} & {\color[HTML]{24292F} 0.6199} & {\color[HTML]{24292F} 0.6199} & {\color[HTML]{24292F} 0.5635} & {\color[HTML]{24292F} 0.8123} & {\color[HTML]{24292F} 0.6731} & {\color[HTML]{24292F} 0.6731} & {\color[HTML]{24292F} 0.6997} & {\color[HTML]{24292F} 0.7158} \\ \hline
{\color[HTML]{24292F} test\_03\_cv\_02} & {\color[HTML]{24292F} 0.6783} & {\color[HTML]{24292F} 0.6783} & {\color[HTML]{24292F} 0.8009} & {\color[HTML]{24292F} 0.6438} & {\color[HTML]{24292F} 0.6169} & {\color[HTML]{24292F} 0.6169} & {\color[HTML]{24292F} 0.8428} & {\color[HTML]{24292F} 0.5326} \\ \hline
{\color[HTML]{24292F} \textbf{mean}} & {\color[HTML]{24292F} 0.6931} & {\color[HTML]{24292F} 0.6931} & {\color[HTML]{24292F} 0.7316} & {\color[HTML]{24292F} 0.7289} & {\color[HTML]{24292F} 0.6342} & {\color[HTML]{24292F} 0.6342} & {\color[HTML]{24292F} 0.6490} & {\color[HTML]{24292F} 0.7176} \\ \hline
{\color[HTML]{24292F} \textbf{std}} & {\color[HTML]{24292F} 0.0447} & {\color[HTML]{24292F} 0.0447} & {\color[HTML]{24292F} 0.0846} & {\color[HTML]{24292F} 0.0606} & {\color[HTML]{24292F} 0.0301} & {\color[HTML]{24292F} 0.0300} & {\color[HTML]{24292F} 0.0762} & {\color[HTML]{24292F} 0.0667} \\ \hline
{\color[HTML]{24292F} \textbf{max}} & {\color[HTML]{24292F} 0.7855} & {\color[HTML]{24292F} 0.7855} & {\color[HTML]{24292F} 0.8318} & {\color[HTML]{24292F} 0.8413} & {\color[HTML]{24292F} 0.6796} & {\color[HTML]{24292F} 0.6790} & {\color[HTML]{24292F} 0.8428} & {\color[HTML]{24292F} 0.7847} \\ \hline
{\color[HTML]{24292F} \textbf{min}} & {\color[HTML]{24292F} 0.6199} & {\color[HTML]{24292F} 0.6199} & {\color[HTML]{24292F} 0.5635} & {\color[HTML]{24292F} 0.6438} & {\color[HTML]{24292F} 0.5638} & {\color[HTML]{24292F} 0.5638} & {\color[HTML]{24292F} 0.5132} & {\color[HTML]{24292F} 0.5326} \\ \hline
\end{tabular}%
}
\end{table}

\section{Scanner differences:}

% \usepackage{graphicx}
\begin{table}[htbp]
\centering
\caption{Architecture used: UNet. Patch size: 128x128 pixels (best fold is reported in bold font). Scanner: Hamamatsu NanoZoomer 2.0-RS.}
\label{tab:my-table5}
\resizebox{0.8\textwidth}{!}{%
\begin{tabular}{|c|c|c|c|c|}
\hline
{\color[HTML]{24292F} \textbf{fold\_name}} & {\color[HTML]{24292F} \textbf{dev\_dice}} & {\color[HTML]{24292F} \textbf{dev\_f1}} & {\color[HTML]{24292F} \textbf{dev\_recall}} & {\color[HTML]{24292F} \textbf{dev\_precision}} \\ \hline
{\color[HTML]{24292F} test\_00\_cv\_00} & {\color[HTML]{24292F} 0.7361} & {\color[HTML]{24292F} 0.7354} & {\color[HTML]{24292F} 0.693} & {\color[HTML]{24292F} 0.8106} \\ \hline
{\color[HTML]{24292F} test\_00\_cv\_01} & {\color[HTML]{24292F} 0.7234} & {\color[HTML]{24292F} 0.7284} & {\color[HTML]{24292F} 0.7451} & {\color[HTML]{24292F} 0.734} \\ \hline
{\color[HTML]{24292F} \textbf{test\_00\_cv\_02}} & {\color[HTML]{24292F} \textbf{0.7388}} & {\color[HTML]{24292F} \textbf{0.7338}} & {\color[HTML]{24292F} \textbf{0.7155}} & {\color[HTML]{24292F} \textbf{0.7744}} \\ \hline
{\color[HTML]{24292F} test\_00\_cv\_03} & {\color[HTML]{24292F} 0.7384} & {\color[HTML]{24292F} 0.732} & {\color[HTML]{24292F} 0.7063} & {\color[HTML]{24292F} 0.7818} \\ \hline
{\color[HTML]{24292F} \textbf{mean}} & {\color[HTML]{24292F} 0.7342} & {\color[HTML]{24292F} 0.7324} & {\color[HTML]{24292F} 0.7150} & {\color[HTML]{24292F} 0.7752} \\ \hline
{\color[HTML]{24292F} \textbf{std}} & {\color[HTML]{24292F} 0.0063} & {\color[HTML]{24292F} 0.0026} & {\color[HTML]{24292F} 0.0191} & {\color[HTML]{24292F} 0.0274} \\ \hline
{\color[HTML]{24292F} \textbf{max}} & {\color[HTML]{24292F} 0.7388} & {\color[HTML]{24292F} 0.7354} & {\color[HTML]{24292F} 0.7451} & {\color[HTML]{24292F} 0.8106} \\ \hline
{\color[HTML]{24292F} \textbf{min}} & {\color[HTML]{24292F} 0.7234} & {\color[HTML]{24292F} 0.7284} & {\color[HTML]{24292F} 0.693} & {\color[HTML]{24292F} 0.734} \\ \hline
\end{tabular}%
}
\end{table}

% \usepackage{graphicx}
\begin{table}[htbp]
\centering
\caption{Architecture used: UNet. Patch size: 128x128 pixels (best fold is reported in bold font). Scanner: Hamamatsu NanoZoomer S60.}
\label{tab:my-table6}
\resizebox{0.8\textwidth}{!}{%
\begin{tabular}{|c|c|c|c|c|}
\hline
{\color[HTML]{24292F} \textbf{fold\_name}} & {\color[HTML]{24292F} \textbf{dev\_dice}} & {\color[HTML]{24292F} \textbf{dev\_f1}} & {\color[HTML]{24292F} \textbf{dev\_recall}} & {\color[HTML]{24292F} \textbf{dev\_precision}} \\ \hline
{\color[HTML]{24292F} test\_00\_cv\_00} & {\color[HTML]{24292F} 0.6286} & {\color[HTML]{24292F} 0.6429} & {\color[HTML]{24292F} 0.6121} & {\color[HTML]{24292F} 0.7143} \\ \hline
{\color[HTML]{24292F} \textbf{test\_00\_cv\_01}} & {\color[HTML]{24292F} \textbf{0.6757}} & {\color[HTML]{24292F} \textbf{0.6695}} & {\color[HTML]{24292F} \textbf{0.6931}} & {\color[HTML]{24292F} \textbf{0.6855}} \\ \hline
{\color[HTML]{24292F} test\_00\_cv\_02} & {\color[HTML]{24292F} 0.617} & {\color[HTML]{24292F} 0.6448} & {\color[HTML]{24292F} 0.6668} & {\color[HTML]{24292F} 0.6443} \\ \hline
{\color[HTML]{24292F} test\_00\_cv\_03} & {\color[HTML]{24292F} 0.6167} & {\color[HTML]{24292F} 0.6445} & {\color[HTML]{24292F} 0.6653} & {\color[HTML]{24292F} 0.6452} \\ \hline
{\color[HTML]{24292F} \textbf{mean}} & {\color[HTML]{24292F} 0.6345} & {\color[HTML]{24292F} 0.6504} & {\color[HTML]{24292F} 0.6593} & {\color[HTML]{24292F} 0.6723} \\ \hline
{\color[HTML]{24292F} \textbf{std}} & {\color[HTML]{24292F} 0.0243} & {\color[HTML]{24292F} 0.0110} & {\color[HTML]{24292F} 0.0294} & {\color[HTML]{24292F} 0.0294} \\ \hline
{\color[HTML]{24292F} \textbf{max}} & {\color[HTML]{24292F} 0.6757} & {\color[HTML]{24292F} 0.6695} & {\color[HTML]{24292F} 0.6931} & {\color[HTML]{24292F} 0.7143} \\ \hline
{\color[HTML]{24292F} \textbf{min}} & {\color[HTML]{24292F} 0.6167} & {\color[HTML]{24292F} 0.6429} & {\color[HTML]{24292F} 0.6121} & {\color[HTML]{24292F} 0.6443} \\ \hline
\end{tabular}%
}
\end{table}